\documentclass[%
 reprint,
 amsmath,amssymb,
 aps,
]{revtex4-2}

\usepackage{graphicx}
\usepackage{dcolumn}
\usepackage{bm}
\usepackage{xcolor}
\usepackage{ulem}

\begin{document}

\preprint{APS/123-QED}

\title{Isotope shift spectroscopy in mercury vapors: a valid alternative to ytterbium for new physics search}

\author{Stefania Gravina}
\affiliation{Dipartimento di Matematica e Fisica, Università degli Studi della Campania ``Luigi Vanvitelli", 81100, Caserta, Italia}
\author{Antonio Castrillo}
\affiliation{Dipartimento di Matematica e Fisica, Università degli Studi della Campania ``Luigi Vanvitelli", 81100, Caserta, Italia}
\author{Livio Gianfrani}%
 \email{livio.gianfrani@unicampania.it}
\affiliation{%
Dipartimento di Matematica e Fisica, Università degli Studi della Campania ``Luigi Vanvitelli", 81100, Caserta, Italia}%

\date{\today}

\begin{abstract}
Precision isotope shift metrology in the deep-UV region has been performed for all bosonic isotopes of mercury with a zero nuclear spin, by using the technique of frequency-comb referenced, wavelength-modulated, saturated absorption spectroscopy. The absolute center frequencies of the 6s$^2$ $^1$S$_0$ $\rightarrow$ 6s6p $^3$P$_1$ intercombination line have been measured with precision in the range of 2.5 - 5.9 10$^{-12}$, in temperature-stabilized mercury vapor samples with natural abundances. Frequency shifts in four isotope pairs have been determined with unprecedented accuracy, the global uncertainty being improved by a factor greater than 20 with respect to the best experimental data of the past literature. Our data set, when combined with previous measurements on the 6s6p $^3$P$_2$$\rightarrow$6s7s $^3$S$_1$ transition at 546 nm, allows us to build a King plot that reveals a nonlinearity with a statistical significance of 4.6$\sigma$.      
\end{abstract}

\maketitle 



Similarly to all alkaline-earth-like metal elements, mercury is a diamagnetic atom with two valence electrons. The energy level diagram, which consists of singlet and triplet states, makes mercury an attractive quantum system for fundamental tests and measurements \cite{Safronova2018}. The lowest triplet states are metastable, the 6s6p $^3$P$_1$ state being connected to the 6s$^2$ $^1$S$_0$ ground state through a relatively narrow intercombination transition, which is normally used for laser cooling and magneto-optical trapping of neutral Hg atoms \cite{Bize2008, Bize2011, Stellmer2022, Bize2023}. Its clock transition (6s$^2$ $^1$S$_0$ $\rightarrow$ 6s6p $^3$P$_0$) is a good candidate for the redefinition of the second based on optical clocks, due to the small sensitivity to blackbody radiation shift \cite{Bize2012}. Furthermore, mercury plays a special role in the history of search for a permanent electric dipole moment (EDM) in diamagnetic atoms \cite{Graner2016}. The fermionic isotope $^{199}$Hg can be used as a magnetometer to monitor magnetic field fluctuations in neutron EDM experiments \cite{Abel2022}. Finally, it is one of the best atomic systems for implementing Doppler broadening thermometry \cite{Truong2015, Gravina2024}. 

Like naturally occurring ytterbium, mercury has seven stable isotopes, five of which are bosons with zero nuclear spin. This latter feature can be exploited for isotope shift spectroscopy, with the ambitious goal of looking for a new physics beyond the Standard Model \cite{Berengut2025}. In fact, a hypothetical new boson that mediates an additional interaction between neutrons and electrons has been shown to provide a contribution to isotope shifts \cite{Delaunay2017, Frugiuele2017}. The latter can be evidenced by using the King-plot analysis method, since it gives rise to King-plot nonlinearity. It is also possible to use the nonlinearity to set an upper bound on a fifth force, thus building the so-called exclusion plots \cite{Berengut2018}. So far, highly significant King-plot nonlinearities have been observed with the analysis of Yb and Yb$^+$ isotope shifts in narrow electric quadrupole transitions \cite{Vuletic2020, Figueroa2022}. Very recently, isotope shift data have been provided for a highly-forbidden electric octupole transition of singly charged ytterbium ions with a sub-kHz precision level \cite{Hur2022, Door2025}. A nonlinearity decomposition plot involving all measured transitions in Yb and Yb$^+$ has shown that the leading cause of nonlinearity is nuclear deformation, the Yb nuclei being prolate deformed \cite{Flambaum2021, Door2025}. However, a second source of nonlinearity cannot be excluded and must be sought among the quadratic field shift and the existence of a new boson \cite{Door2025}. 

The even-even stable isotopes of mercury, with Z=80, have only two protons less than the magic number 82 and are known to have a nearly spherical nuclear shape, approaching that of the doubly magic $^{208}$Pb nucleus. Calculations of the quadrupole deformation parameter ($\beta_2$) by means of the finite-range droplet model (FRDM) have found relatively small values, ranging from -0.052 for $^{204}$Hg to -0.125 for the rare $^{196}$Hg isotope \cite{Moller2016}. The same calculations for the even-even stable isotopes of ytterbium yield $\beta_2$ values between 0.286 and 0.300 \cite{Moller2016}. Consequently, nuclear deformation contributions to the isotope shift in mercury even-even isotopes are expected to be much smaller than those of Yb isotopes.

In this paper, we report on a thorough experimental investigation of the isotope shift of the Hg intercombination transition (6s$^2$ $^1$S$_0$ $\rightarrow$ 6s6p $^3$P$_1$) at 253.7 nm. Absolute center frequencies are measured for the five bosonic isotopes ($^{196}$Hg, $^{198}$Hg, $^{200}$Hg, $^{202}$Hg and $^{204}$Hg) with an unprecedented accuracy by using the technique of wavelength modulated, comb-locked, Lamb-dip spectroscopy \cite{Gianfrani2024}. Data analysis reveals King-plot nonlinearity with a statistical significance of 4.6$\sigma$. 

To a good approximation, the shift in frequency of an atomic transition $i$ for a given isotope pair (A, A$^\prime$) can be expressed as the sum of two terms: the mass shift, coming from the nuclear recoil, and the field shift, which is due to the penetration of the electronic wavefunction into the finite nuclear charge distribution. Each term can be factored into a transition-dependent electronic component and a nuclear component that depends on the isotope pair, so that the following equation holds \cite{King1984}:   
\begin{equation}
\label{eq:isotopeshift}
\nu_{i}^{A,A^\prime} = \nu_{i}^{A} - \nu_{i}^{A^\prime} = K_i \mu^{A,A^\prime}+F_i \delta\!<\!r^2\!>_{A,A^\prime}.    
\end{equation}
Here $\delta\!<\!r^2\!>_{A,A^\prime} = <\!r^2\!>_{A}-<\!r^2\!>_{A^\prime}$ is the variation in the mean-square nuclear charge radius, $\mu^{A,A^\prime}=1/M_A-1/M_{A^\prime}$ is the change in the inverse nuclear mass, $K_i$ and $F_i$ are electronic coefficients quantifying the mass shift and the field shift, respectively. Combining the mass-scaled isotope shifts of a pair of transitions, $1$ and $2$, for the same isotope sequence, it is possible to eliminate $\delta\!<\!r^2\!>_{A,A^\prime}$ and build a linear equation that is the basis of a King plot \cite{King1984}. More particularly, the linearity is expressed by the equation reported hereafter:   
\begin{equation}
\label{eq:Kingplot}
\tilde\nu_{2}^{A,A^\prime} = \frac{F_2}{F_1}\tilde\nu_{1}^{A,A^\prime}+K_2-K_1\frac{F_2}{F_1},    
\end{equation}
where $\tilde\nu_{i}^{A,A^\prime}$ is the mass-scaled isotope shift, given by $\nu_{i}^{A,A^\prime}/\mu^{A,A^\prime}$. Additional terms in Eq. (\ref{eq:isotopeshift}) should be considered to take into account higher-order contributions, such as the quadratic field shift and nuclear deformation variation, given by $G_i^{(2)}\delta\!<\!r^2\!>_{A,A^\prime}^2$ and $G_i^{(4)}\delta\!<r^4>_{A,A^\prime}$, respectively \cite{Flambaum2018}.  
Similarly, a further correction to Eq. (\ref{eq:isotopeshift}) might be due to an attractive force between neutrons of the atomic nucleus and bound electrons, mediated by a hypothetical new boson \cite{Frugiuele2017}. More particularly, it has been shown that, if the interaction is modeled by a Yukawa-like potential, the additional term can be written as $\alpha_{NP}X_ih^{A,A^\prime}$, thus having the same characteristic as the other terms, being factored into electronic and nuclear components \cite{Frugiuele2017, Berengut2018}. In this expression, $\alpha_{NP}$ quantifies the coupling strength of the new boson to electrons and neutrons, $h^{A,A^\prime}$ is simply given by $A-A^\prime$, namely, the difference in neutron number between the two isotopes, while $X_i$ gives the effect of the Yukawa potential on the atomic transition $i$.\\
Therefore, the isotope shift becomes:
\begin{multline}
    \label{eq:fullisotopeshift}
\nu_{i}^{A,A^\prime} = K_i \mu^{A,A^\prime}+F_i \delta\!<\!r^2\!>_{A,A^\prime}+G_i^{(2)}\delta<\!r^2\!>_{A,A^\prime}^2+\\+G_i^{(4)}\delta<\!r^4\!>_{A,A^\prime}+\alpha_{NP}X_ih^{A,A^\prime}. 
\end{multline}
The addition of the last three terms makes the linearity of the King plot no longer strictly valid. It should be noted that there is another source of nonlinearity that can be ascribed to the nuclear polarizability effect, a circumstance that adds further complexity to the interpretation of possible King plot nonlinearities \cite{Flambaum2018}.

The experimental set-up is an upgraded version of that described in \cite{Gianfrani2024}. The main improvements are outlined hereafter, while further details can be found in the supplemental material. Firstly, we used a longer mercury vapor cell. This latter consists of a 2-cm-long quartz tube, sealed at the two ends by a pair of wedged windows with a diameter of 25.4 mm. The cell is equipped with a small reservoir in the middle of its length. As guaranteed by the manufacturer, the cell is baked at 425 °C for a minimum of 24 h in an ultrahigh vacuum environment, prior to the injection of liquid mercury in the reservoir. This procedure ensures that ultrapure Hg vapors are in equilibrium with the liquid phase inside the cell. The system is placed in a thermostatic vacuum chamber, whose details can be found in \cite{Lopardo2021}. So doing, it is possible to actively control the temperature in the range between $\sim$220 and $\sim$320 K, with a sub-millikelvin stability for several hours. These modifications allowed us to lower the Hg pressure by a factor of $\sim$10, as compared to \cite{Gianfrani2024}, thus reducing the width of the sub-Doppler profile as well as the uncertainty in the retrieval of the line center frequency. In this regard, an exception was made necessary for $^{196}$Hg line. In fact, due to the small abundance of the $^{196}$Hg isotope ($\sim$0.15$\%$), the temperature was set to about 301 K so that the mercury atomic density was sufficiently high to recover the sub-Doppler line with a satisfactory signal-to-noise ratio (SNR).   
The chamber is magnetically shielded by a mu-metal layer in order to minimize the interaction of the mercury vapors with the Earth’s magnetic field.

\begin{figure*}[htbp]
\includegraphics[width=0.44\linewidth]{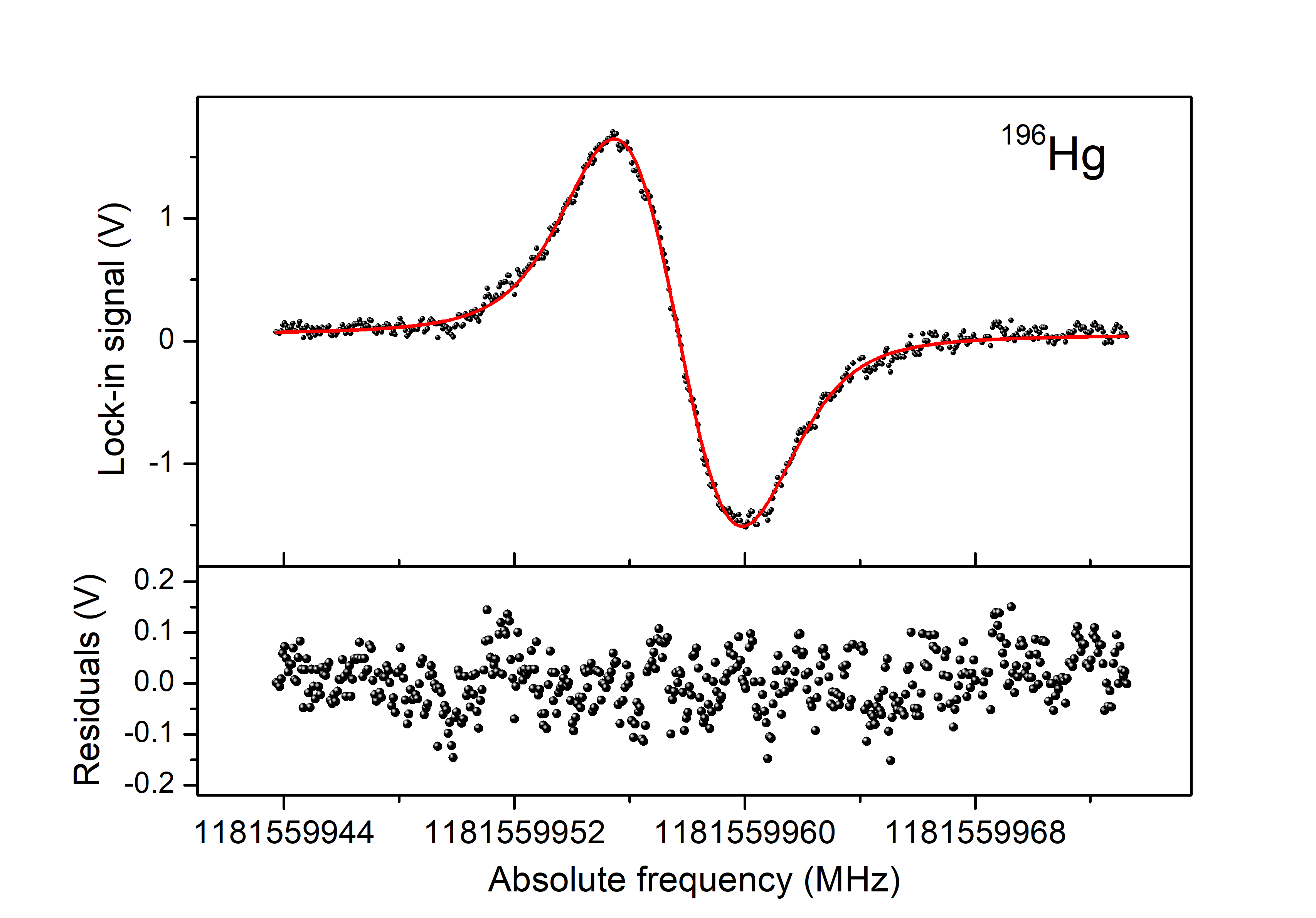}
\quad{\includegraphics[width=0.44\linewidth]{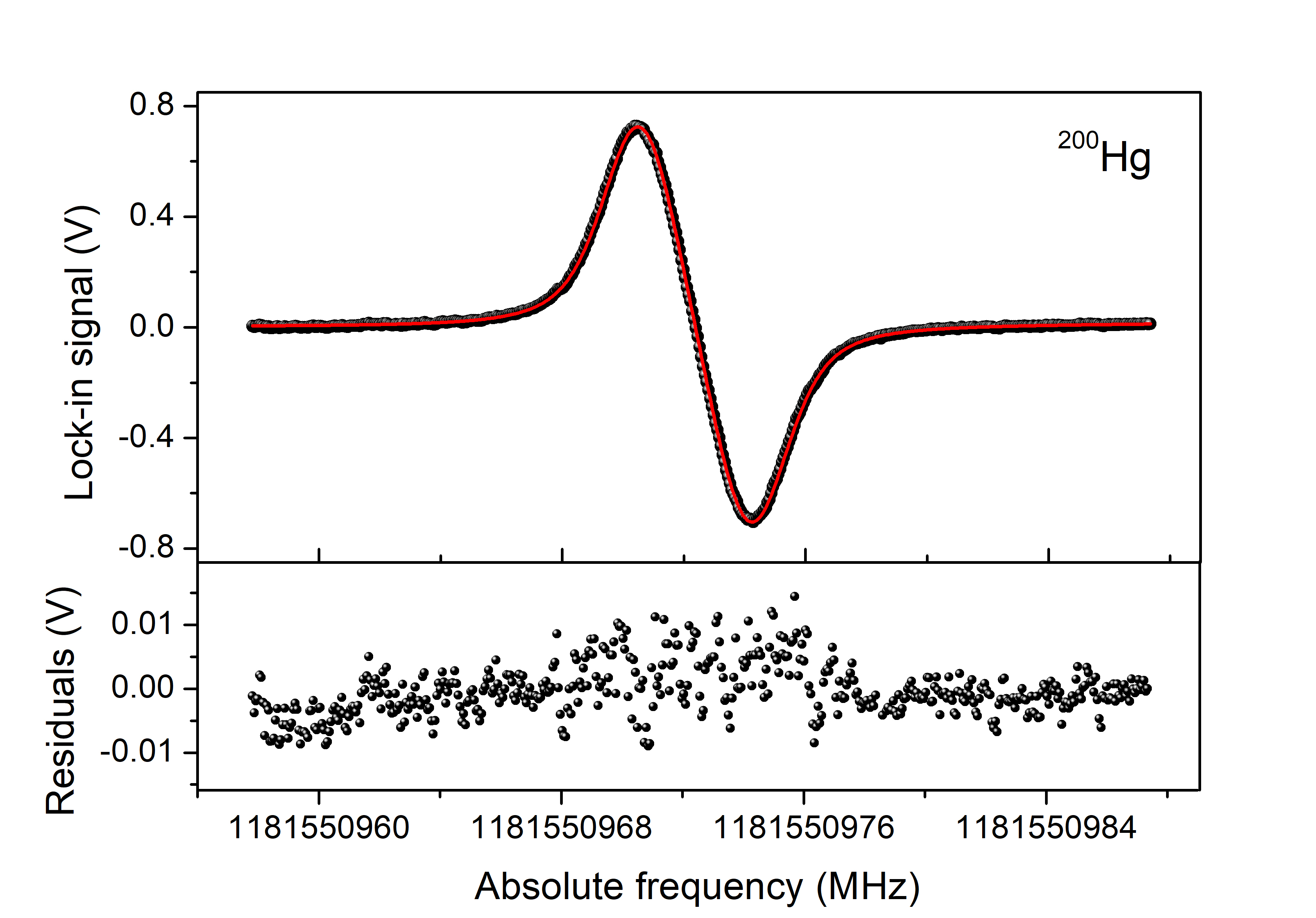}}
\caption{\label{spectra} Examples of sub-Doppler spectra for the $^{196}$Hg (left panel) and $^{200}$Hg (right panel) intercombination transition. The red lines represent the best-fit curve of experimental data to the wavelength-modulated Lamb-dip profile. The bottom plot of both panels shows the residuals. The signal-to-noise ratio amounts to $\sim$60 and $\sim$360 for $^{196}$Hg and $^{200}$Hg, respectively. }
\end{figure*}

The optical layout is identical to that of \cite{Gianfrani2024}. Very briefly, we recall that the deep UV radiation is produced through a double stage of second harmonic generation in a pair of nonlinear crystals, starting from the near-infrared (NIR) wavelength of 1014.8 nm, as emitted by an external cavity diode laser (ECDL) \cite{Clivati2020}. The NIR radiation is locked to the nearest tooth of an optical frequency comb synthesizer (OFCS), which in turn is referenced to a GPS-disciplined Rb clock \cite{Gravina2022}. A wavelength-modulation technique was implemented for measuring nonlinear spectroscopic signals in a pump-probe configuration, with an incident power of $\sim$3 $\mu$W. To this aim, the offset frequency between the ECDL and the nearest comb-tooth was dithered at 200 Hz with a modulation index of $\sim$1. This dither propagates throughout the frequency chain, very effectively, from the NIR to the deep-UV region. Comb-calibrated UV frequency scans around an arbitrary center frequency were carried out by finely tuning the comb repetition rate. First-harmonic synchronous
detection of the probe beam power was performed by means of an analog lock-in amplifier, with an integration time of 100 ms. Particular care was taken in setting the saturation degree to the optimal value (between 1 and 2) for the largest Lamb-dip contrast. For this reason, the waist size of the pump beam was monitored using a sensitive deep-ultraviolet camera, while the power was measured by means of a highly precise optical power meter. Example spectra are shown in Figure \ref{spectra}.
Line fitting was performed using the wavelength modulated Voigt function, as described in \cite{Axner2012}.
The fit residuals demonstrate an
excellent agreement between the experimental data and the line shape model, which is an important prerequisite for the accurate retrieval of the line center frequencies. It is worth noting that the spectral analysis gives a full width at half maximum (FWHM) of the sub-Doppler line of about 3 MHz, with the only exception of the $^{196}$Hg line, for which a 5-MHz wide profile was found. Therefore, a significant reduction in the linewidth was observed, compared to the measurements of \cite{Gianfrani2024}, as a result of the improvements described above. 

The experimental results are summarized in Table~\ref{tab:table1}. The uncertainties in absolute frequencies are of pure statistical nature and are calculated as 1$\sigma$ standard deviation of the mean values resulting from 20 repeated determinations for each mercury isotope, as recorded at a constant value of the incident intensity. No correction was applied to the line center frequencies to account for the ac-Stark effect and the pressure shift. 
These corrections are reported in the last two columns of Table~\ref{tab:table1}, while the intensity values of the pump beam are given in the 6$^{th}$ column. In this framework, the $^{200}$Hg isotope makes an exception. In fact, only for this isotope, Lamb-dip spectroscopy was performed for different values of the pump-beam intensity, so that it was possible to extrapolate the zero-intensity line center frequency and the ac-Stark shift per unit intensity. The latter was found to be -2.23 (12) kHz$/$mW$/$cm$^2$. This value shows a satisfactory agreement with the determination reported in \cite{Gianfrani2024}, the relative uncertainty being a factor of $\sim$ 4 smaller than previously found (for further details, see the supplemental material).
For the aims of pressure shift corrections, the estimated coefficient of the Hg-Hg collision-induced frequency shift is -33 (11) kHz$/$Pa \cite{Witkowski2019}. The joint occurrence of these two effects is considered in the isotope shift data and contributes to the associated uncertainty budget.  
\begin{figure}[htbp]
\centering
\includegraphics[width=0.9\linewidth]{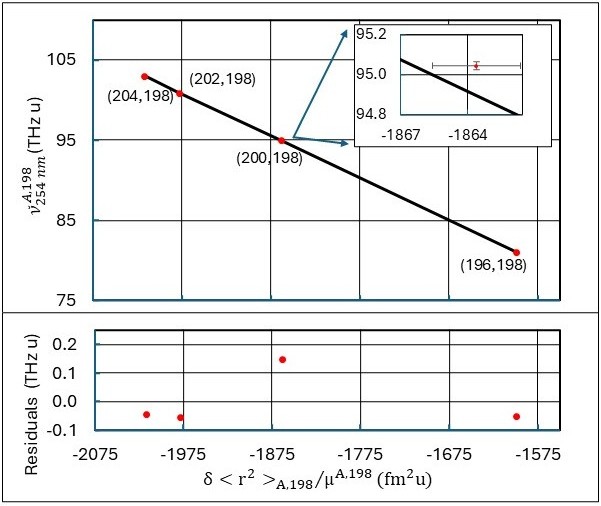}
\caption{Mass-scaled isotope shifts for the Hg intercombination transition versus the differences in the mean-square nuclear charge radii divided by the proper inverse-mass difference. The best-fit line is drawn in red, while the absolute residuals are reported in the bottom plot. The inset graph shows the (200,198) data point for which the greatest deviation from the best fit line is observed. The vertical error bar is multiplied by a factor of 100 to be displayed in the graph.}
\label{figure2}
\end{figure}
Table~\ref{tab:table2} summarizes the isotope shift determinations that result from absolute frequency variations with respect to the reference isotope $^{198}$Hg. We also report the atomic masses (expressed in atomic mass units, u) \cite{Wang2021}, the inverse-mass differences ($\mu^{A,198}$), and the differences in mean-square nuclear charge radii ($\delta\!<\!r^2\!>_{A,198}$) \cite{Angeli2013}.
Our data are compared with theoretical values based on multiconfiguration Dirac-Hartree-Fock calculations
with configuration interaction \cite{McFerran2022}, and the best experimental values from the past literature \cite{Witkowski2019}. 
The uncertainty budget for the isotope shift of each pair is given in Table~\ref{tab:table3}. In addition to the entries due to the uncertainty in the ac-Stark shift and pressure shift, we considered the frequency calibration as a further source of type-B (systematic) uncertainty. Specifically, the OFCS introduces an uncertainty of $\sim$3 kHz to each absolute frequency measurement in the deep-UV domain. This is due to the relative stability of the GPS-disciplined Rb clock, which is 2.5 10$^{-12}$ at 1 s. Other sources of uncertainty, including the Zeeman shift due to the residual magnetic field into the cell, gas lens, and wavefront-curvature-induced residual Doppler effect, are quoted to be negligible. In addition, residual gases that may form in the sealed cell, such as helium, water, and molecular hydrogen, are expected to act in the same way on all Hg isotopes. Therefore, possible pressure-induced perturbations cancel out. Returning to the discussion of Table~\ref{tab:table2}, a satisfactory agreement occurs between our values and those of \cite{Witkowski2019}, while significant deviations are found from the theoretical values of \cite{McFerran2022}. 

\begin{table*}[t]
\caption{\label{tab:table1}%
Measured values of the absolute center frequency of the Hg intercombination line for all bosonic isotopes.  
}
\begin{ruledtabular}
\begin{tabular}{llllllll}
\textrm{A} &
\textrm{Abundance\footnote{Data taken from Ref. \cite{Manfred2016}.}} &
\textrm{Temperature} &
\textrm{Pressure\footnote{Details about pressure determinations are in the Supplemental Material.}} &
\textrm{Frequency} &
\textrm{Intensity} &
\textrm{ac-Stark shift} &
\textrm{Pressure shift}\\
\textrm{} &
\textrm{} &
\textrm{(K)} &
\textrm{(Pa)} &
\textrm{(kHz)} &
\textrm{(mW/cm$^2$)} &
\textrm{(kHz)} &
\textrm{(kHz)}\\

\colrule
196 & 0.0015(1) & 301.15 & 0.334(3) &1181 559 957 711(6) & 7.5&-16.7(9) &-11(4) 
\\
198 & 0.1004(3) & 273.65 & 0.0284(3) &1181 555 777 703(5) & 50.1&-112(6) &-0.9(3)
\\
200 & 0.2314(9) & 273.65 & 0.0284(3)&1181 550 972 297(3)  &0 &0 &-0.9(3)
\\
202 & 0.2974(13) & 273.65 & 0.0284(3)&1181 545 676 641(6)  &20.1 &-45(2) &-0.9(3)
\\
204 & 0.0682(4) & 273.65 & 0.0284(3)&1181 540 466 132(7)  &57.4 &-128(7) &-0.9(3)
\end{tabular}
\end{ruledtabular}
\end{table*}

\begin{table*}[htbp]
\caption{\label{tab:table2}%
Isotope shift determinations compared with theoretical values \cite{McFerran2022} and the best experimental data of past works \cite{Witkowski2019}. The uncertainties reported in the 5$^{th}$ column include statistical and systematic components, as explained in Table~\ref{tab:table3}.  
}
\begin{ruledtabular}
\begin{tabular}{ccccccc}
\textrm{A} &
\textrm{Atomic mass} &
\textrm{$\mu^{A,198}$} &
\textrm{$\delta\!<\!r^2\!>_{A,198}$} &
\textrm{$\nu_{254}^{A,198}$}&
\textrm{$\nu_{254}^{A,198}$}&
\textrm{$\nu_{254}^{A,198}$}\\
 & (u) & (10$^{-5}$ u$^{-1})$ & $(10^{-3}fm^2)$ &(kHz)& (kHz)&(kHz)\\
 &Ref. \cite{Wang2021} & & Ref. \cite{Angeli2013} & This work& Ref. \cite{McFerran2022}& Ref. \cite{Witkowski2019}

\\
\colrule
196 & 195.965833(3) & 5.157754(8)
 & -82.5(0.1) &4 179 923(11)& 4 406 000&-

\\
198 & 197.9667692(5) & 0 & 0 &0&0&0
\\
200 & 199.9683269(6) & -5.056088(2)
 & 94.2(0.1)&-4 805 518 (9)& - 5 054 000&-4 806 280 (330)
\\
202 & 201.9706436(8) & -10.013823(2)
 & 198.1(0.1)&-10 101 129 (11)& -10 579 000&-10 101 740 (280)
\\
204 & 203.9734940(5) & -14.875505(2)
 & 300.1(0.1)&-15 311 555 (13)& -16 060 000&-15 312 020 (300)

\end{tabular}
\end{ruledtabular}
\end{table*}

\begin{table*}[t!]
\caption{\label{tab:table3}%
Uncertainty budget for the isotope shift data. All values are in kHz. The overall uncertainties are used to set the error bars of figures \ref{figure2} and \ref{figure3}.}
\begin{ruledtabular}
\begin{tabular}{llllllll}
\textrm{Source}&
\textrm{Type}&
\textrm{$\nu_{254}^{196,198}$}&
\textrm{$\nu_{254}^{200,198}$}&
\textrm{$\nu_{254}^{202,198}$}&
\textrm{$\nu_{254}^{204,198}$}&
\textrm{$\nu_{254}^{200,202}$}&
\textrm{$\nu_{254}^{202,204}$}\\
\hline
Statistics&A& 7.8&5.8 &7.8 &8.6& 6.7&9.2 \\
Pressure& & & & & & &\\
shift &B & 4.0&0.42 &0.42 &0.42 &0.42 &0.42 \\
ac-Stark& & & & & & &\\
shift&B & 6.1 &6.0 &6.3 &9.2 &2.0 &7.3\\
Frequency& & & & & & &\\
calibration&B &4.2 & 4.2& 4.2& 4.2&4.2 &4.2\\
\hline
Overall&A+B &11 &9.4 &11&13 & 8.2& 12\\
\end{tabular}
\end{ruledtabular}
\end{table*}

\begin{figure}[htbp]
\centering
\includegraphics[width=1\linewidth]{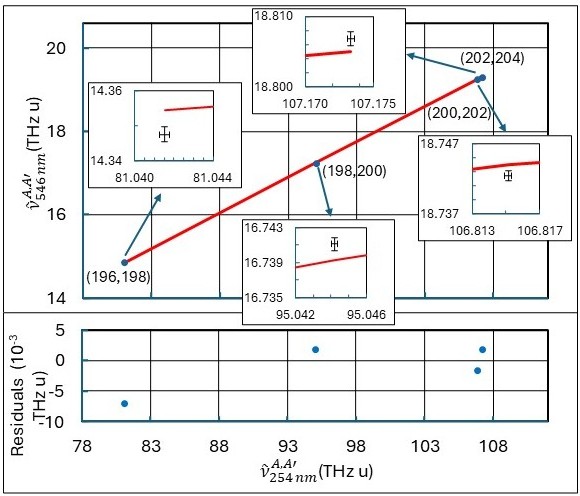}
\caption{King plot of the Hg 6$^3$P$_2\to$7$^3$S$_1$ transition at 546 nm (data from Ref. \cite{Hall1989}) and 6$^3$S$_1\to$6$^3$P$_1$ intercombination transition at 254 nm. The numbers in parentheses next to the data points indicate the mass numbers of the isotope pair used to calculate the modified isotopic shift between the two transitions. The error bars on both the x- and y-axes correspond to 1$\sigma$. A deviation from linearity (red line) of 4.6$\sigma$ is observed. The residuals (in the lower panel) exhibits a zigzag pattern of the type -+-+ \cite{Vuletic2020}.}
\label{figure3}
\end{figure}

Our data set provides useful information on the electronic coefficients, $K_{254}$ and $F_{254}$, for the mercury intercombination transition.
In Figure \ref{figure2}, the mass-scaled isotope shifts are plotted as a function of $\delta\!<\!r^2\!>_{A,A^\prime}/\mu^{A,A^\prime}$, using the variation in the mean-square nuclear charge radius provided by \cite{Angeli2013}. A weighted linear fit of these data (taking into account the uncertainties of both variables)  yields the values of $K_{254}$ and $F_{254}$, as readily derived from Eq. (\ref{eq:isotopeshift}).
The retrieved values are $K_{254}$=(-2.67 $\pm$ 0.47) THz u and $F_{254}$=(-52.37 $\pm$ 0.24) GHz/fm$^2$.  
It should be noted that there is a good linearity for our data set, as demonstrated by the Pearson's coefficient, which is equal to -0.99996. However, this is not sufficient to exclude possible deviations from linearity, despite the relatively large uncertainties on the mean-square nuclear charge radii variations. In this regard, suspicion falls on the point (200,198), which is close to a possible disagreement, as clearly shown in the inset of Figure \ref{figure2}.
To better investigate this issue, it is useful to resort to the King plot formalism, which allows one to eliminate the dependence on $\delta\!<\!r^2\!>_{A,A^\prime}$. Building a King-plot requires isotope shift data for at least two spectral lines, as shown in Eq. \ref{eq:Kingplot}. To this end, we considered the 
6s6p $^3$P$_2$$\rightarrow$6s7s $^3$S$_1$ transition at 546 nm, whose isotope shift data are taken from \cite{Hall1989}.
In Figure \ref{figure3}, the mass-scaled isotope shifts of the 546-nm transition are plotted against the data of the intercombination line. The relative uncertainties on $\tilde{\nu}_{254~ nm}^{A,A'}$ range from 1.6$\times$10$^{-6}$ to 3.1$\times$ 10$^{-6}$, while those on $\tilde{\nu}_{546~nm}^{A,A'}$ vary between 3.6$\times$10$^{-5}$ and 1.4$\times$10$^{-4}$ for the same isotope pairs. Also in this case, we performed a weighted linear fit taking into account the uncertainties on both variables. The absolute residuals are shown in the bottom plot of Figure \ref{figure3}. 
It should be noted that the analysis of a King plot requires particular care due to the existence of correlations between the data points \cite{Vuletic2020}. A generalised-least-squares (GLS) method should be used, a procedure that needs as input the covariance matrix of the data \cite{Hur2022}. This is not an issue, as long as the covariance matrix is positive definite. Complications arise if the matrix is negative definite, like the one we built for the isotope shift data of the 546-nm transition. 
However, it is possible to retrieve information about the amount of correlation by using the Durbin-Watson statistic test \cite{Durbin1950}. The test statistic ranges from 0 to 4, with values close to 2 indicating no autocorrelation in the residuals of a regression analysis. In our case, we found a value of 1.73 that justifies the approach of disregarding the off-diagonal elements of the covariance matrix.

The slope of the best-fit line yields the ratio $F_{546}/F_{254} = 0.17031(6)$, while the intercept is determined to be 0.552(6) THz u. The normalized $\chi^2$ of the regression is 26, which corresponds to a p-value of 2.3$\times 10^{-6}$ and a nonlinearity significance of 4.6$\sigma$. 
The slope is consistent with the less accurate value 0.17047 (11) of \cite{Witkowski2019}.

We performed precision measurements of the absolute frequency of the mercury intercombination line at 253.7 nm for all bosonic isotopes, including the rare $^{196}$Hg isotope. An upgraded version of the spectrometer described in \cite{Gianfrani2024}, based on the technique of comb-locked wavelength-modulated saturated absorption spectroscopy, allowed us to determine the isotope shifts of four isotope pairs, improving the accuracy by more than a factor of 20 compared to the best determinations of the past literature \cite{Witkowski2019}. 
By combining our results with existing data for the 546-nm transition, we constructed a King plot that reveals a nonlinearity with a statistical significance of 4.6$\sigma$. This is a remarkable result that requires confirmation with new and improved measurements on the transition at 546 nm.
After this step, it would be important to carry out an in-depth analysis of the King-plot nonlinearity, looking for a possible signature of a new physics beyond the Standard Model. In this regard, because of its nuclear structure, mercury appears to be a very promising atomic system with some advantages compared to ytterbium, as discussed in this letter.
\begin{acknowledgments}
We wish to acknowledge the support of the Italian Ministry for Education and Research (MUR) within the PRIN 2022 program through the MENPHYS project (Grant n. 2022KMNSR9-01).
\\
\end{acknowledgments}

\nocite{*}

\providecommand{\noopsort}[1]{}\providecommand{\singleletter}[1]{#1}%

\end{document}